\documentclass[useAMS,usenatbib]{mn2e}

\usepackage{psfig}

\title[A periodic variation in Ser\,X-1]{A 2 hour periodic variation in the low mass X-ray binary Ser\,X-1}

\author[R. Cornelisse et~al.]{R. Cornelisse$^{1,2}$\thanks{E-mail:
    corneli@iac.es}, J. Casares$^{1,2}$, P.A. Charles$^{3,4}$, D. Steeghs$^{5}$\\
  $^{1}$ Instituto de Astrofisica de Canarias, Via Lactea, La Laguna, 
E-38200, Santa  Cruz de Tenerife, Spain\\
  $^2$ Departamento de Astrofisica, Universidad de La Laguna, 
E-38205, La Laguna, Tenerife, Spain\\
  $^3$ School of Physics and Astronomy, University of Southampton, 
Highfield, Southampton SO17 1BJ, UK\\
  $^4$ Department of Astronomy, University of Capetown, Private Bag X3, 
Rondebosch 7701, South Africa\\
  $^5$ Department of Physics, University of Warwick, Coventry, CV4 7AL, UK\\
}
\begin{document}

\date{Accepted Received; in original form}

\pagerange{\pageref{firstpage}--\pageref{lastpage}} \pubyear{2013}

\maketitle

\label{firstpage}

\begin{abstract}
Spectroscopy of the low mass X-ray binary Ser\,X-1 using the Gran
Telescopio Canarias have revealed a $\simeq$2 hr periodic variability
that is present in the three strongest emission lines. We tentatively
interpret this variability as due to orbital motion, making it the
first indication of the orbital period of Ser\,X-1. Together with the
fact that the emission lines are remarkably narrow, but still
resolved, we show that a main sequence K-dwarf together with a
canonical 1.4$M_\odot$ neutron star gives a good description of the
system. In this scenario the most likely place for the emission lines
to arise is the accretion disk, instead of a localized region in the
binary (such as the irradiated surface or the stream-impact point),
and their narrowness is due instead to the low inclination
($\le$10$^{\circ}$) of Ser\,X-1.
\end{abstract}


\section{Introduction}

Low mass X-ray binaries (LMXBs) are interacting binaries where a
low-mass donor transfers matter onto the neutron star or black hole
via Roche-lobe overflow. This process forms an accretion disk and
gives rise to the observed X-rays, making them among the brightest
X-ray sources in the sky (see e.g. the collected reviews in Lewin \&
van der Klis 2006). Although optical counterparts for many of the
bright LMXBs have been known for a long time, kinematic studies of
these systems are often not straightforward. This is mainly due to
reprocessing of the X-rays in the outer accretion disk that completely
dominates the optical light. This situation has changed only with the
advent of sensitive spectrographs on very large telescopes.

Using phase-resolved medium resolution spectroscopy of Sco\,X-1,
Steeghs \& Casares (2002) detected the presence of narrow high
excitation lines that were strongest in the Bowen region (a blend of
N\,III $\lambda$4634/4640 and C\,III $\lambda$4647/4650). These narrow
lines originate from the irradiated surface of the donor star and
allowed for the first time a radial velocity study of both binary
components in Sco\,X-1 which led to constrain its mass
function. Phase-resolved spectroscopic studies of about a dozen other
bright LMXBs have revealed the presence of these narrow components in
the Bowen region in most LMXBs and provided in many cases the first
constraints on the mass of their compact objects (e.g. Cornelisse
et~al. 2008 for an overview).

One such bright LMXB is Serpens\,X-1 (Ser\,X-1). It was discovered in
the 1960s (Friedman et~al. 1967), but its optical counterpart
(MM\,Ser) was only identified about 10 years later (Davidsen
1975). Further observations taken in good seeing conditions showed
that the optical counterpart consisted of a blend of two stars that
are separated by $\simeq$2.1 arcsec (Thorstensen
et~al. 1980). Subsequently, one of these was itself revealed to be a
blend of two stars that are only separated by 1 arcsec (Wachter
1997). Both spectroscopy (Hynes et~al. 2004) and radio observations
(Migliari et~al. 2004) confirmed that the bluer of the stars is the
true optical counterpart. Due to the well-established presence of
thermonuclear X-ray bursts from Ser\,X-1, the compact object is known
to be a neutron star (Li \& Lewin 1976). Furthermore, Ser\,X-1 was one
of the first neutron star LMXBs in which a broad relativistic iron
line was detected, and modeling of this line inferred an inclination
of $<$25$^{\circ}$ (Bhattacharyya \& Strohmayer 2007; Ng
et~al. 2010). However, despite over 40 years of study little else is
known about Ser\,X-1.

\begin{figure*}\begin{center}
\psfig{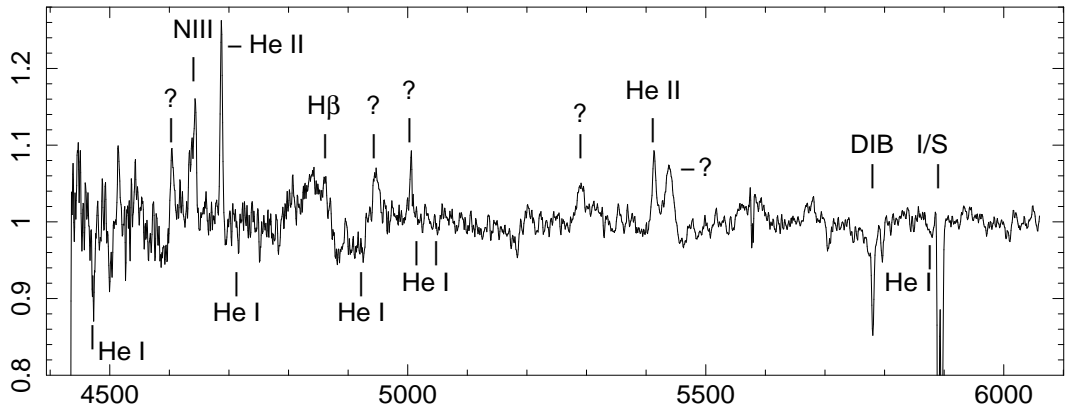}
\caption{Average normalized OSIRIS/GTC spectrum of Ser\,X-1. Indicated
  are the most prominent identified emission lines, sky lines (I/S)
  and diffuse interstellar bands (DIB) while the unidentified features
  are labeled with ``?''. We note that several of these unidentified
  features were previously observed by Hynes et~al. (2004).
\label{spect}}
\end{center}\end{figure*}

Here we present the results of a spectroscopic study of Ser\,X-1 using
the Gran Telescopio Canarias (GTC) at the Observatorio del Roque de
los Muchachos on the Canary Island of La Palma (Spain).  We detect a
$\simeq$2 hr periodic variation that we interpret as the orbital
period. Although the emission lines are very narrow and sharp (as was
already reported by Hynes et~al. 2004), we present further evidence
that this is due to the low inclination of the system instead of the
lines being formed on the irradiated surface of the donor star or the
stream-impact point.

\section{Observations and Data Reduction}

On 6 July 2011 we observed Ser\,X-1 for 4 hrs continuously (UT
21:45-02:23) using OSIRIS on GTC. For each exposure we used the
volume-phased holographic grism R1200V with a slit width of 1.0
arcsec, resulting in a total of 24 spectra with an exposure time of
600 seconds each. The seeing during the observations varied between
0.7 and 1.0 arc-seconds. Since Ser\,X-1 is $\simeq$1 arcsec from a
fainter field star, we aligned the slit in such a way that we
minimized the contribution of this interloper. However, this does mean
that no comparison star is included in the slit and we could therefore
not correct for slit losses. Arc lamp exposures were taken for
wavelength calibration during the daytime.

For the data reduction (i.e. debias, flatfielding etc) we used the
PAMELA software that allows for optimal extraction of the spectra
(Horne 1986). We determined the pixel-to-wavelength scale using a 4th
order polynomial fit to 20 lines resulting in a dispersion of 0.79
\AA\,pixel$^{-1}$, an rms scatter of $<$0.05 \AA, and a wavelength
coverage of $\lambda$$\lambda$4434-6060. The final resolution around
the He\,II $\lambda$4686 emission line is 107.6 km s$^{-1}$
(FWHM). Finally we also corrected for any velocity shifts due to
instrument flexure (always $<$6 km s$^{-1}$) by cross-correlating the
sky spectra. For the corresponding analysis of the resulting dataset
we used the MOLLY package.
 
\section{Data Analysis}

\subsection{Spectrum}

\begin{figure}\begin{center}
\psfig{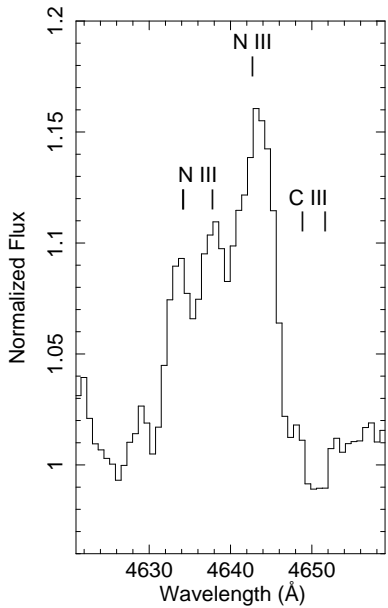}
\end{center}
\caption{Close-up of the Bowen blend of Ser\,X-1 with the position of
  the expected N\,III and C\,III lines indicated. Note that the
  spectrum has not been corrected for the systemic velocity of
  $\simeq$92 km s$^{-1}$.
\label{bowen}}
\end{figure}

First we created an average normalized spectrum of Ser\,X-1, and show
the result in Fig.\,\ref{spect}. The spectrum is very similar to that
presented by Hynes et~al. (2004), but there are also a few
differences. To start with the similarities, the most prominent
emission lines that are commonly detected in LMXBs are also present in
our spectrum (and indicated in Fig.\,\ref{spect}), and all of them are
extremely narrow as shown in Table\,\ref{stats}. In particular a
close-up of the Bowen region, presented in Fig.\,\ref{bowen}, suggests
that the individual N\,III components (i.e. 4634 and 4640\AA) are
clearly separated, while there is also no hint of any C\,III lines.
Furthermore, we detect He\,I $\lambda$4471 in absorption, and there
are hints of features at the location of other He\,I transitions in
our spectrum (which are indicated in Fig.\,\ref{spect}).  However,
since He\,I $\lambda$4471 is on the edge of the wavelength coverage
where the sensitivity starts to drop significantly, it is not possible
to conclude anything else from this line. Finally, just like Hynes
et~al. (2004) we also detect the broad absorption trough just redward
of H$\beta$ and the unidentified emission features at 4604 and 4945
\AA. This confirms beyond doubt that they are not artefacts and must
be true features of the system, and we have indicated them with a
``?'' in Fig.\,\ref{spect}.

The two main differences between our spectrum and that of Hynes
et~al. (2004) are a much more complex structure around H$\beta$ and
several new features (which we have also indicated with a ``?''). The
detailed structure around H$\beta$ (ranging from 4800-4950 \AA)
appears to form an inverted P Cygni profile, as if it were a single
feature. However, within the current framework of LMXBs it is
difficult to understand how a significant infall component above the
disk of Ser\,X-1 at velocities well over 4000 km s$^{-1}$ (which are
needed to produce such a profile) could occur. Furthermore, from the
He\,I $\lambda$4471 absorption line we already know that at least
He\,I $\lambda$4922 will be present in absorption and contributing to
this feature. We therefore think it more likely that the full complex
around H$\beta$ is made up of many blended emission and absorption
lines.

Of the new features that are present in our spectrum, the broad
emission components just redward of He\,II $\lambda$5411 and the
narrow one around 5004 \AA, are particularly interesting. They should
have been clearly detected by Hynes et~al. (2004) and to our knowledge
have no counterpart in any other LMXB. To identify the narrow features
at 4605 and 5004 \AA, we used the atomic line list by van Hoof \&
Verner (1998) and the applied off-set of 92 km s$^{-1}$ (see
Sect. 3.2) from the He\,II $\lambda$4686 rest-frame wavelength as an
estimate of the systemic velocity of Ser\,X-1. The 4605 \AA\, feature
corresponds to a N\,III transition (4603.8 \AA), while that at 5004
\AA\, is close to a N\,II one (5002.7 \AA). We therefore tentatively
conclude that these unknown transitions are due to ionised Nitrogen.

We examined the broad feature at 5440\AA\, in the individual spectra
by eye and note that it is extremely variable. Although present in all
spectra, its strength and width shows large changes from spectrum to
spectrum. We illustrate this in Fig.\,\ref{variable} by showing the
close-up of this region in two different spectra. Although we have no
explanation for the strong variability displayed, the profile in the
lower spectrum in Fig.\,\ref{variable} does suggest that it consists
of several unresolved lines. Indeed we note that a large number of
N\,II/N\,III transitions occur in this region that are close to the 
individual peaks displayed, but the quality and resolution of the 
individual spectra does not allow us to unambiguously conclude this.

\subsection{Radial velocities}

\begin{figure}
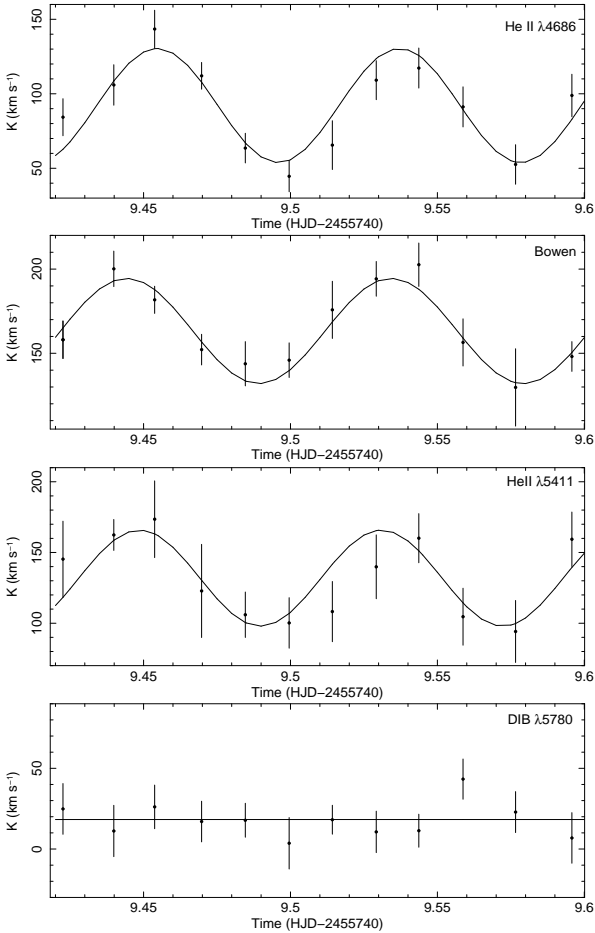
\begin{center}
\psfig{figure=rv_heii.ps,angle=-90,width=8cm}
\psfig{figure=rv_bowen.ps,angle=-90,width=8cm}
\psfig{figure=rv_he5411.ps,angle=-90,width=8cm}
\psfig{figure=rv_dib.ps,angle=-90,width=8cm}
\end{center}
\caption{Radial velocity curves for the 3 strongest emission lines and
  a diffuse interstellar band (around 5780\AA) present in the spectrum
  of Ser\,X-1. The emission lines are from top to bottom: He\,II
  $\lambda$4686, N\,III $\lambda$4640 and He\,II
  $\lambda$5411. Overplotted is the best fitting sine curve for each
  line. The radial velocity curve of the diffuse interstellar band on
  the other hand shows no obvious velocity movement.
\label{rv}}
\end{figure}

Following the report of Hynes et~al. (2004) of a small drift of He\,II
$\lambda$4686 over the period of $\simeq$an hour, we inspected the
trail of this line by eye. We could clearly see some movement over
time and therefore averaged two consecutive spectra together to
improve the signal-to-noise, thereby creating 12 spectra in total. For
each of these spectra we then determined the full width at
half-maximum (FWHM), equivalent width (EW) and radial velocity. Our
FWHM measurements show that the line is resolved and we therefore
corrected our measurements for instrumental broadening using a
nearby line in our arc spectrum. Since both the FWHM and EW do not
change significantly during our observations we present their average
value in Table\,\ref{stats}. Finally, we note that the average radial
velocity is consistent with that observed by Hynes et~al. (2004), and
in Fig.\,\ref{rv} we show the resulting radial velocity curve.

\begin{figure}
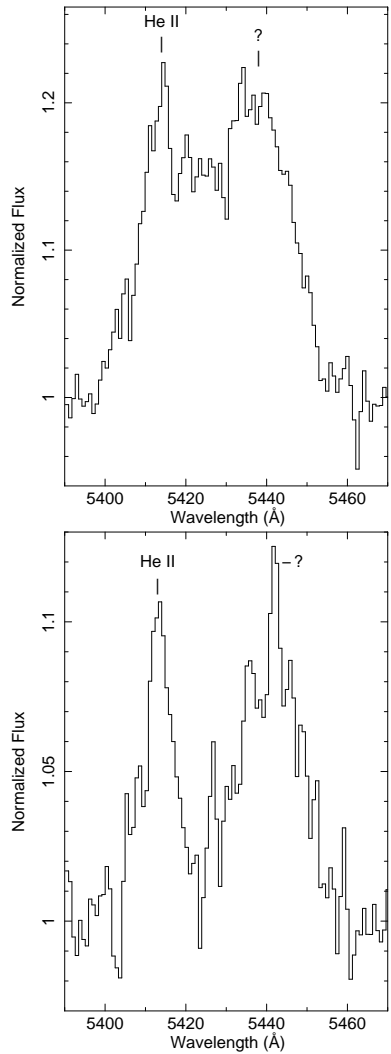
\begin{center}
\psfig{figure=He5411_unclear.ps,angle=-90,width=5cm}
\psfig{figure=He5411_clear.ps,angle=-90,width=5cm}
\end{center}
\caption{Close-up of the He\,II $\lambda$5411 emission line for two
  spectra of Ser\,X-1. Although He\,II is clearly detected in both
  spectra (and labeled as such), a broad and highly variable
  unidentified feature is also present (labeled with ``?'').
\label{variable}}
\end{figure}

\begin{table}\begin{center}
\caption{Emission line properties in Ser\,X-1 for both lines that are
  typically observed in LMXBs (known lines) and emission lines that
  are unique to Ser\,X-1 (unknown lines). Included are the best fit
  period ($P$) and semi-amplitude ($K$) obtained via sine-wave fits
  for the lines where variability is observed. For the lines that are
  part of a blend we have also indicated the properties of the full
  region. All FWHM measurements are the intrinsic broadening after
  deconvolving the instrumental resolution.
\label{stats}}
\begin{tabular}{lcccc}
\hline
 Line  & EW  & FWHM & $P$ & $K$\\
       & (\AA)   & (km s$^{-1}$) & (hrs) & (km s$^{-1}$)\\
\hline
\multicolumn{5}{l}{\bf known lines}\\
He\,II $\lambda$4686 & 1.20$\pm$0.07 & 180$\pm$10 & 1.98$\pm$0.13 & 38$\pm$5\\
N\,III $\lambda$4640 & 0.81$\pm$0.06 & 210$\pm$12 & 2.19$\pm$0.15 & 31$\pm$5\\
He\,II $\lambda$5411$^{1}$ & 1.01$\pm$0.04 & 305$\pm$44 & 1.98$\pm$0.20 & 34$\pm$8\\
\hline
\multicolumn{5}{l}{\bf unidentified lines}\\
$\lambda$4604 & 0.68$\pm$0.08 &  418$\pm$62\\
$\lambda$4945 & 0.57$\pm$0.06 & 1061$\pm$94\\
$\lambda$5004 & 0.30$\pm$0.04 &  246$\pm$34 & 1.84$\pm$0.34$^2$ & 23$\pm$12$^2$\\
$\lambda$5290 & 0.42$\pm$0.05 &  823$\pm$85\\
\hline
{\bf blends}\\
Bowen                & 1.40$\pm$0.09 & 717$\pm$57   & &\\
He\,II $\lambda$5411$^3$ & 3.62$\pm$0.08 & 2220$\pm$138 & &\\
\hline
\multicolumn{5}{l}{$^1$EW and FWHM are for the He\,II emission line only.}\\
\multicolumn{5}{l}{$^2$Note that the sine-wave fit is formally not
  significant.}\\
\multicolumn{5}{l}{$^3$EW and FWHM are for the full structure between 5400-5450\AA}\\
\end{tabular}
\end{center}\end{table}

Interestingly, the radial velocity curve does appear to show periodic
motion. We therefore checked other strong emission lines (i.e. the
Bowen region and He\,II $\lambda$5411) to see if similar variability
is present. The Bowen region also shows that there are no strong
changes between the spectra in both strength and width of the
individual components that make up the blend. We also note that there
are clear velocity shifts in N\,III $\lambda$4640, the strongest
component of the Bowen region, and therefore measured the FWHM, EW and
radial velocity for this line. We present the results in
Fig.\,\ref{rv} and Table\,\ref{stats}. Finally we also measured the
average EW and FWHM of the full Bowen blend, and list these
results in Table\,\ref{stats}.

As already pointed out in Sect.\,3.1, the region around He\,II
$\lambda$5411 is more complex due to the strongly variable component
just redward (see Fig.\,\ref{variable}). However, He\,II is itself
clearly visible in the individual spectra and again it is moving, and
in Fig\,\ref{rv} we give its radial velocity curve. However, the FWHM
and in particular EW measurements of the individual spectra gives such
a large spread (see Fig.\,\ref{variable}) that we have opted to only
measure these values from the average of all spectra, giving the
results in Table\,\ref{stats}. Finally we also measured the FWHM and
EW of the full complex between 5400-5450 \AA, and these are also
included in Table\,\ref{stats}.

We also searched for variability in the unidentified emission lines in
Fig.\,\ref{spect} (those indicated with ``?''). Most lines are either
too weak or show no variability, and we therefore only present their
FWHM and EW in Table\,\ref{stats}. The emission line at 5004 \AA\ is
the only one that shows some evidence of variability although at a low
significance of $\simeq$2$\sigma$. Interestingly, a sine-wave fit to
its radial velocity curve gives a period and amplitude that are
consistent with the lines discussed above. For completeness we list
the fit results in Table\,\ref{stats}, but will not include them in
any further analysis.

Fig\,\ref{rv} shows that the three emission lines all have a
variability with a period, semi-amplitude and phasing that are similar
within the errors. To verify that the variability does not have an
instrumental origin we also measured the radial velocities of a
diffuse interstellar band (DIB) around 5780 \AA. This is included as
the bottom panel of Fig\,\ref{rv}, and shows clearly no velocity
shifts in the DIB feature to a limit of $<$9.4 km s$^{-1}$ (99\%
confidence). Finally, for the He\,II$\lambda$4686 radial velocity
curve a $\chi$$^2$-test indicates that the variability is highly
significant, since fitting it to a constant velocity gives
$\chi^2_\nu$=5.7 (for 11 degrees of freedom).

We therefore conclude that the variability in the 3 emission lines is
real, and fitted the radial velocity curve for each emission line with
a sine wave (resulting in $\chi^2_\nu$=0.85 for He\,II $\lambda$4686).
We list the resulting periods and semi-amplitudes in
Table\,\ref{stats}, and note that apart from the systemic velocity,
$\gamma$, all parameters are similar within the errors. The
differences in $\gamma$ are most likely due to the fact that both
N\,III $\lambda$4640 and He\,II $\lambda$5411 are blended with nearby
lines, thereby producing an off-set. This is further strengthened by
the fact that their FWHM is much broader than He\,II
$\lambda$4686. Since we do not expect He\,II $\lambda$4686 to be
blended we think that this line provides the best estimate of the
systemic velocity, which we measure as $\gamma$=92$\pm$4 km s$^{-1}$.

Finally, we phase-folded both the Bowen blend and He\,II $\lambda$4686
in 8 bins to check for other emission features moving around on a
similar period (but at a different phase and semi-amplitude). A visual
inspection does not show any other moving component. Furthermore, we
created averages by co-adding spectra spanning 0.25 orbital cycles
around the expected minimum and maximum of both N\,III $\lambda$4640
and He\,II $\lambda$4686. At both extrema the lines can be very well
described by a single Gaussian and the addition of a second Gaussian
(either as a broader underlying feature or a second narrow peak) does
not significantly improve the fit (according to an $F$-test). We
therefore conclude that the emission lines can be represented by a
single component.

\section{Discussion}

We have presented evidence for a $\simeq$2 hrs periodic radial
velocity variation in the emission lines from Ser\,X-1 from our GTC
observations. Although this variation is based only on approximately 2
cycles worth of data, and therefore still needs confirmation, we
tentatively conclude that we have detected the orbital period of
Ser\,X-1. For the rest of the discussion we will assume that the
variability is truly due to orbital motion and consider what the
implications are for the properties of this system.

Another important observation is the fact that most emission lines are
narrow and are best described by a single component. This strongly
suggests that they are located in a single region somewhere in the
binary. Hynes et~al. (2004) have already listed several potential
locations: the outer accretion disk, the irradiated surface of the
donor star or the stream-impact point. Based on our proposed orbital
period, we can now explore these different suggestions to find out
what the consequences for the system parameters of Ser\,X-1 will be.

Could the emission lines arise from the irradiated surface of the
donor star? Narrow components arising from the irradiated donor star
surface have been observed in many persistent LMXBs (see
e.g. Cornelisse et~al. 2008 for an overview). These narrow components
are in general strongest in the Bowen region, but there are also
examples (e.g. EXO\,0748$-$676) where the He\,II lines show a large
component arising from the irradiated surface (Mu\~noz-Darias
et~al. 2009).  Although the narrow components in many LMXBs have a
FWHM comparable to that observed for Ser\,X-1, their radial velocity
semi-amplitudes are typically an order of magnitude larger
(e.g. Casares et~al. 2006, Cornelisse et~al. 2007, Barnes
et~al. 2007). If the lines arise from the irradiated surface of the
donor star, then the observed FWHM of the emission lines of
$\simeq$180 km s$^{-1}$ must be due to rotational broadening only
(since there is no evidence that the lines consist of multiple
components). In this case we can easily show that for any reasonable
mass ratio of Ser\,X-1, the FWHM is incompatible with the observed
radial velocity semi-amplitude of $K_{\rm em}$=38 km s$^{-1}$.

As a first order estimate to interpret our measurements, we must
assume a large mass ratio and that the accretion disk in Ser\,X-1 is
not shadowing the donor star, i.e. $K_{\rm em}$ is tracing emission
coming from L1.  This will lead to the largest so-called
$K$-correction in order to derive the true radial velocity
semi-amplitude of the donor star. Using the $K$-correction polynomials
by Mu\~noz-Darias et~al. (2005) for $q$=0.8 (the largest mass ratio
for which their $K$-correction is still valid to 1\%) gives a radial
velocity semi-amplitude of the donor star of $K_2$=111 km
s$^{-1}$. From this value of $K_2$, together with $q$=0.8 we can
estimate a rotational broadening of the emission lines of $V$sin$i$=
0.462$K_{2}$$q^{1/3}$(1+$q$)$^{2/3}$=70 km s$^{-1}$ (Wade \& Horne
1988).  This is much smaller than the 180 km s$^{-1}$ we observe, and
suggests that $q$$\gg$0.8. Even for the highly unlikely assumption of
$q$$\simeq$1.0 these numbers will still not work, and we cannot obtain
a sensible solution, and we therefore reject the suggestion that the
lines arise from the donor star.

If we instead assume that the emission lines are produced in the
accretion disk (or the stream-impact region) we find that it is
possible to create a consistent set of system parameters. As an X-ray
burster, the compact object is a neutron star, whose mass we will
assume is close to the canonical value of 1.4$M_\odot$. Furthermore,
from our proposed short orbital period we can also infer that the
donor star, if it is a main sequence star, must be of a late spectral
type such as a M5V dwarf (which has a mass of 0.14$M_\odot$). Under
these assumptions we can use Kepler's law to calculate that the binary
radius is $a$=0.9$R_\odot$. Furthermore the donor's Roche lobe would
then be 0.2$R_\odot$ (Paczynski 1971), i.e. comparable to the radius
of such a late type star. 

Interestingly, de Jong et~al. (1996) derived a mean disk opening angle
for LMXBs of $\simeq$12$^\circ$, making it possible that the disk in
Ser\,X-1 is indeed flared sufficiently to shield the donor star
completely from irradiation. This could explain the absence of any
indication of emission from the irradiated donor star in Ser\,X-1. On
the other hand, the presence of ionised Nitrogen, combined with the
absence of Carbon suggest a more massive donor star. Ser\,X-1 does not
show any hint of the typically observed C\,III $\lambda$4647/4650 in
its spectrum (see e.g. Steeghs \& Casares 2002), while the curious
emission features in Fig.\,\ref{spect} (those labelled with ``?'')
could all be Nitrogen transitions. This strange abundance could be due
to the CNO fusion reaction cycle (e.g. Shen \& Bildsten 2007), where
proton capture on $^{14}$N is the slowest step. Since the CNO cycle is
only dominant in stars more massive than our sun, this would imply
that the donor in Ser\,X-1 was originally much more massive and that
the processed material was mixed to the surface by
convection. However, we know from Hynes et~al. (2004) that H$\alpha$
has a similar strength as is typically observed in LMXBs (note that
from our Fig.\,\ref{spect} it is not even clear if H$\beta$ is
detected), suggesting that the outer layers of the donor in Ser\,X-1
have not been stripped. It is therefore unclear if the spectrum of
Ser\,X-1 can account for an evolved low mass donor.

Despite the caveat mentioned above we can still consider whether the
emission is coming from the full outer accretion disk or a localized
spot in the disk such as the stream-impact region using a low mass
donor. In the first case the observed radial velocity curve should
trace the motion of the compact object, while in the second case it
would be that of the Keplerian orbit of the impact region. In both
scenarios we need an estimate of the size of the accretion disk, and
we assume that it reaches at least the truncation radius ($R_{\rm
  trunc}$). Using the system parameters described above we estimate
that $R_{\rm trunc}$=0.60$a$/(1+$q$)=0.45$R_\odot$ (Frank
et~al. 2002). Assuming a Keplerian accretion disk, the velocity at the
edge of the disk would then be $\simeq$770 km s$^{-1}$. If the line
emission arises in the stream-impact region, which we expect to be
close to the edge of the disk, an inclination of
$i$$\simeq$3$^{\circ}$ is needed to explain the observed 38 km
s$^{-1}$ semi-amplitude. Furthermore, this inclination would also
imply a non-projected FWHM of $\simeq$3650 km s$^{-1}$. Since these
numbers, in particular the true FWHM, are rather extreme we think this
unlikely.

For the case of disk emission on the other hand, we can estimate an
inclination of the system of $i$$\simeq$9$^{\circ}$ using the mass
function
($f$($M_2$)=$M_2$sin$^3i$/(1+$q$)$^2$=$K_1^3$$P$/2$\pi$$G$). This
suggests that the emission lines would have a non-projected FWHM of
$\simeq$1150 km s$^{-1}$, which is compatible to the FWHMs observed in
high-inclination LMXBs such as X1822$-$371 or GR\,Mus (Casares
et~al. 2003; Barnes et~al. 2007). Furthermore, if Ser\,X-1 is observed
at such low inclination the spectral resolution of our dataset would
not allow us to resolve the double-peaked structure that is typical
for emission lines that are formed in the accretion disk (and have
typical peak separations of $\simeq$600 km s$^{-1}$), and the emission
lines become indistinguishable from a single gaussian.

Obviously, more complicated system geometries can be envisaged. One
such complication, which is most likely important in Ser\,X-1, is the
presence of strong tidal forces. For example Whitehurst (1988) showed
that the accretion disk becomes tidally unstable and asymmetric for
systems with an extreme mass-ratio (i.e. $q$$\le$0.3). Since this is
most likely the case in Ser\,X-1 the accretion disk will not be a
simple Keplerian one as we have assumed here. However, we do not think
that this will affect our main conclusion, namely that we have
discovered a $\simeq$2 hr periodic variability that we tentatively
identify as the orbital period. If true, Ser\,X-1 is a low inclination
system with the spectral lines dominated by emission from the
accretion disk.

\section*{Acknowledgments}
This work is based on data collected at the Observatorio del Roque de
los Muchachos, La Palma, Spain [Obs.Id. GTC7-11A]. We would like to
thank the referee, Craig Heinke, for helpful comments which have
improved this paper. We acknowledge the use of PAMELA and MOLLY which
were developed by T.R.  Marsh, and the use of the on-line atomic line
list at http://www.pa.uky.edu/$\sim$peter/atomic.  RC acknowledges a
Ramon y Cajal fellowship (RYC-2007-01046) and a Marie Curie European
Reintegration Grant (PERG04-GA-2008-239142). RC and JC acknowledge
support by the Spanish Ministry of Science and Innovation (MICINN)
under the grant AYA 2010-18080. This program is also partially funded
by the Spanish MICINN under the consolider-ingenio 2010 program grant
CSD 2006-00070. DS acknowledges support from STFC.

\bsp

\label{lastpage}

\end{document}